\documentclass{emulateapj}

\newcommand{\Lx}{\ensuremath{L_{\mathrm{X}}}}
\newcommand{\Yx}{\ensuremath{Y_{\mathrm{X}}}}

\newcommand{\Msol}{\ensuremath{\mathrm{M_{\odot}}}}

\newcommand{\rt}{\ensuremath{R_{\mathrm{200}}}}
\newcommand{\rf}{\ensuremath{R_{\mathrm{500}}}}

\newcommand{\Mf}{\ensuremath{M_{\mathrm{500}}}}
\newcommand{\Mt}{\ensuremath{M_{\mathrm{200}}}}

\newcommand{\Zsol}{\ensuremath{\mathrm{Z_{\odot}}}}

\newcommand{\Mgas}{\ensuremath{M_{\mathrm{gas}}}}

\newcommand{\OM}{\ensuremath{\Omega_{\mathrm{M}}}}

\newcommand{\eg}{{\it e.g.\ }}
\newcommand{\egc}{{\it e.g.}}  

\newcommand{\ie}{{\it i.e.\ }}

\newcommand{\Chandra}{\emph{Chandra}}
\newcommand{\Einstein}{\emph{Einstein}}
\newcommand{\ROSAT}{\emph{ROSAT}}
\newcommand{\XMM}{\emph{XMM-Newton}}

\newcommand{\chisq}{\ensuremath{\chi^2}}

\newcommand{\gta}{\,\rlap{\raise 0.4ex\hbox{$>$}}{\lower 0.6ex\hbox{$\sim$}}\,}  
\newcommand{\lta}{\,\rlap{\raise 0.4ex\hbox{$<$}}{\lower 0.6ex\hbox{$\sim$}}\,}  

\newcommand{\nm}{\mbox{\ensuremath{\mathrm{~\nm}}}}

\newcommand{\km}{\mbox{\ensuremath{\mathrm{~km}}}}

\newcommand{\kpc}{\mbox{\ensuremath{\mathrm{~kpc}}}}
\newcommand{\Mpc}{\mbox{\ensuremath{\mathrm{~Mpc}}}}
\newcommand{\s}{\mbox{\ensuremath{\mathrm{~s}}}}

\newcommand{\keV}{\mbox{\ensuremath{\mathrm{~keV}}}}
\newcommand{\erg}{\mbox{\ensuremath{\mathrm{~erg}}}}

\newcommand{\pMpc}{\ensuremath{\mathrm{\Mpc^{-1}}}}
\newcommand{\ps}{\ensuremath{\mathrm{\s^{-1}}}}

\newcommand{\ergps}{\ensuremath{\mathrm{\erg \ps}}}

\newcommand{\kmpspMpc}{\ensuremath{\mathrm{\km \ps \pMpc\,}}}
\newcommand{\LT}{\mbox{\ensuremath{\mathrm{L_{X}-kT}}}}

\newcommand{\LY}{\mbox{\ensuremath{\mathrm{L_{X}-Y_{X}}}}}
\newcommand{\YM}{\mbox{\ensuremath{\mathrm{Y_{X}-M_{500}}}}}
\newcommand{\LM}{\mbox{\ensuremath{\mathrm{L_X-M}}}}
\newcommand{\LMf}{\mbox{\ensuremath{\mathrm{L_X-M_{500}}}}}

\newcommand{\MgL}{\mbox{\ensuremath{\mathrm{M_{gas}-L_{X}}}}}

\newcommand{\MT}{\mbox{\ensuremath{\mathrm{M-kT}}}}

\newcommand{\LCDM}{$\Lambda$CDM~}

\begin{document}

\title{The \LY\ relation: Using galaxy cluster X-ray luminosity as a robust, low scatter mass proxy.}

\author{B. J. Maughan\altaffilmark{1}}
\affil{Harvard-Smithsonian Center for Astrophysics, 60 Garden St, Cambridge, MA 02140, USA.}
\altaffiltext{1}{\Chandra\ fellow}
\email{bmaughan@cfa.harvard.edu}

\shorttitle{The \LY\ relation.}
\shortauthors{B. J Maughan}


\begin{abstract}
We use a sample of 115 galaxy clusters at $0.1<z<1.3$ observed with
\Chandra\ ACIS-I to investigate the relation between luminosity and \Yx\
(the product of gas mass and temperature). The scatter in the relation is
dominated by cluster cores, and a tight \LY\ relation ($11\%$ intrinsic
scatter in \Lx) is recovered if sufficiently large core regions ($0.15\rf$)
are excluded. The intrinsic scatter is well described by a lognormal
distribution and the relations are consistent for relaxed and
disturbed/merging clusters. We investigate the \LY\ relation in low-quality
data (e.g. for clusters detected in X-ray survey data) by estimating \Lx\
from soft band count rates, and find that the scatter increases somewhat to
$21\%$. We confirm the tight correlation between \Yx\ and mass and the
self-similar evolution of that scaling relation out to $z=0.6$ for a subset
of clusters in our sample with mass estimates from the literature. This is
used to estimate masses for the entire sample and hence measure the \LM\
relation. We find that the scatter in the \LM\ relation is much lower than
previous estimates, due to the full removal of cluster cores and more
robust mass estimates. For high-redshift clusters the scatter in the \LM\
relation remains low if cluster cores are not excluded. These results
suggest that cluster masses can be reliably estimated from simple
luminosity measurements in low quality data where direct mass estimates, or
measurements of \Yx\ are not possible. This has important applications in
the estimation of cosmological parameters from X-ray cluster surveys.
\end{abstract}


\section{Introduction}
The number density of clusters of galaxies and the details of their growth
from the highest density perturbations in the early Universe are sensitive
to the underlying cosmology. The high X-ray luminosities of clusters makes
it relatively easy to detect and study clusters to high redshifts with
X-ray telescopes. For this reason, X-ray cluster studies have been
effectively used to impose tight constraints on various cosmological
parameters \citep[e.g.][]{hen00,vik03,all04}. Many cluster catalogs have
been assembled based on clusters detected by \Einstein\
\citep[e.g.][]{gio90a} and \ROSAT\ \citep[e.g.][]{sch97,ebe98,vik98b,boh01}
with further surveys planned and underway with \Chandra\ and \XMM\
\citep[e.g.][]{rom01}. These provide large statistically complete samples
of clusters that are ideal for cosmological studies.

However, the challenge in using clusters as cosmological probes lies in
obtaining reliable mass estimates for the large numbers of clusters
required for statistical studies. The most reliable X-ray mass estimates
for individual clusters are obtained by solving the equation of hydrostatic
equilibrium, which requires measurements of the density and temperature
gradients of the X-ray emitting gas. This is possible for nearby, bright
clusters, but the majority of clusters detected in surveys are too faint
for such detailed observations. In these cases, cluster masses must be
estimated from simple properties such as X-ray luminosity (\Lx) or a single
global temperature (kT). Furthermore, statistical samples include many
disturbed clusters for which the assumption of hydrostatic equilibrium is
undermined, causing additional difficulties in deriving their masses.

Self-similar models, in which the only significant energy source in
clusters is their gravitational collapse, predict simple scaling relations
between basic cluster properties and the total mass. Observational
studies have found that such scaling relations exist,
but that their form is not identical to the self-similar predictions. For
example, the slope of the X-ray luminosity-temperature (\LT) relation is
steeper than self-similar predictions \citep[\egc][]{mar98a,arn99} and the
entropy in cluster cores is higher than predicted
\citep[\egc][]{pon99,pon03}. This indicates the importance of
non-gravitational effects (such as cooling, mergers, and jets from AGN) on
the energy budget of clusters. While the form of the 
scaling relations differs from self-similarity, it is generally found that
the evolution of the scaling relations follows the self-similar predictions
\citep[e.g.][]{vik02,mau06a}. This suggests that the evolution is dominated
by the changing density of the Universe.

In addition to the shape and evolution of the scaling relations, a key
consideration is their scatter. Not only are low-scatter scaling relations
desirable for obtaining more precise mass estimates, but the scatter must
be understood in order to account for the effects of bias in samples that
are defined based on an observable property (e.g. \Lx) that has some finite
intrinsic (i.e. not due to measurement errors) scatter with mass. The
scatter in the scaling relations has been found to be dominated by the
cluster cores \citep[\egc][]{oha06,che07}. The high gas density in many cluster cores
leads to rapid radiative cooling of the gas which condenses and is replaced
by gas cooling and flowing in from larger radii \citep{fab94b}. This
runaway cooling is thought to be balanced by energy input from AGN or
supernovae, but leads to bright central peaks in gas density profiles and
low core temperatures. Such cool-core clusters will then deviate from
self-similar scaling relations involving \Lx\ and/or kT, though these
effects can be reduced by correcting for the cool-core emission by
excluding core regions from the measurements \citep[e.g.][]{mar98a} or
including an additional parameter to model the effect \citep{oha06,che07}. A
second important source of scatter is cluster mergers, which can cause
transient increases in \Lx\ and kT
\citep{ran02}. This can cause clusters to scatter along the \LT\ relation,
but away from the mass-temperature (\MT) relation \citep{row04,kra06a}. A
further consideration is that the effects of cool cores and mergers have
different redshift dependences. At $z\ga0.5$ the frequency of cool cores is
lower, and of merging clusters is higher, than in the local universe
\citep[][]{jel05,vik06c,mau07b}.

The X-ray emissivity of cluster gas is proportional to the square of its
density, so the processes discussed above can have strong effects on the
luminosity. Indeed, the luminosity-mass (\LM) relation, while not commonly
studied, has been found to have a large intrinsic scatter
\citep{rei02}. The \MT\ relation, meanwhile has been studied in detail and
the scatter is found to be considerably smaller
\citep[e.g.][]{fin01,san03,arn05,vik06a}. Finally, a cluster's gas mass
(\Mgas) is believed to be simply related to its total mass by a universal
baryon fraction \citep[e.g.][]{all04}. Both \Mgas\ and kT are popularly
used as proxies for the total mass in cosmological studies
\citep[e.g.][]{vik03,hen04}. Recently \citet[][hereafter
KVN06]{kra06a} used simulations and observations to show that the parameter
\Yx (the product of the X-ray temperature and gas mass) is a superior mass
proxy to either quantity alone, exhibiting a low-scatter scaling relation
with mass, and being insensitive to cluster mergers. This result has been
verified independently by the simulations of \citet{poo07}.

While \Yx\ is an excellent mass proxy, the \LM\ relation remains of
fundamental importance both for estimating masses when data quality only
permit luminosity measurements, and for characterising the biases in
cosmological studies based on X-ray flux-limited samples. In this {\it
paper} we use a large sample of clusters observed with
\Chandra\ to study the \LY\ relation for the first time, and by using \Yx\
as a mass proxy, we investigate the \LM\ relation. A \LCDM cosmology of
$H_0=70\kmpspMpc\equiv100h\kmpspMpc$, and $\OM=0.3$
($\Omega_\Lambda=0.7$) is adopted throughout. All errors are quoted at the
$68\%$ level.

\section{Sample and data analysis}
The sample consists of all galaxy clusters observed with ACIS-I in the
\Chandra\ public archive as of November 2006 with published redshifts greater than 0.1; $115$
clusters at $0.103<z<1.26$ (34 at $z>0.5$). The construction and analysis of the sample is
discussed in detail in
\citet{mau07a}, but we repeat some pertinent points here. Gas masses were
obtained by projecting the 3D emissivity profile of \citet[][hereafter V06]{vik06a} along the
line of sight, and fitting this to the projected emissivity profile
computed from the surface brightness distribution of each
cluster. Temperatures were measured by fitting an absorbed, single
temperature APEC \citep{smi01} model to the spectrum extracted from the
aperture $(0.15<r<1)\rf$. Luminosities were either measured from a spectral
fit in the aperture of interest, or for the purposes of computing values
comparable to those measured in survey data, from the count rate within
that aperture (see \textsection \ref{s.surv}). The lower redshift cutoff of
the sample ensures that \rf\ falls within the ACIS-I field of view for all
clusters, avoiding any extrapolation in the \Lx\ measurements.

Following the method outlined by KVN06, the radius \rf\ was determined
iteratively, measuring the temperature in the aperture $(0.15<r<1)\rf$ and
the gas mass within \rf, computing a new \Yx, and hence estimating a new
value of \rf. In order to estimate \rf\ from \Yx\ we used the \YM\ relation
measured for the \citet{vik06a} sample of clusters
\begin{eqnarray}\label{e.ym}
M_{500} & = & \frac{h}{0.72}^{\frac{5B_{YM}}{2}-1}C_{YM}E(z)^{a_{YM}}\frac{Y_X}{6\times10^{14}\Msol\keV}^{B_{YM}},
\end{eqnarray}
with $B_{YM}=0.564$, $C_{YM}=7.047\times10^{14}\Msol$ and
$a_{YM}=-2/5$ (A. Vikhlinin, priv. comm.). The
evolution of the relation depends on $E(z) = [\OM(1+z)^3 +
(1-\OM-\Omega_\Lambda)(1+z)^2 + \Omega_\Lambda]^{1/2}$.  This observed \YM\
relation has a normalisation $\sim15\%$ lower than that found in the
simulations of KVN06, possibly due to the effect of turbulent pressure
support which is neglected in the mass derivations for the observed
clusters (KVN06). Adopting the normalisation from the simulations would
have the effect of increasing \Lx\ by a small amount, as \rf\ would
increase by $\sqrt[3]{15}\%$. The normalisation of the \LM\ relations
derived in \textsection \ref{sec:estim-mass-from} also would increase by
$15\%$.

For 4 of the faintest clusters (RXJ0910+5422, CLJ1216+2633,
CLJ1334+5031 and RXJ1350.0+6007), temperatures could not be constrained
when the central $0.15\rf$ was excluded. For those clusters, all of the
cluster emission ($r<\rf$) was used for the temperature measurements in
measuring \rf\ and \Yx. Core regions were excluded for the luminosity
measurements as required. The measured properties of all of the clusters
are given in \citet{mau07b}.

\section{The \LY\ relation}\label{s.yl}
The self-similar \LY\ relation can be obtained by a simple combination of
the \MgL\ and \LT\ relations to give
\begin{eqnarray}\label{e.yl}
L_X & = & C_{LY}E(z)^{a_{LY}}\frac{Y_X}{4\times10^{14}\Msol\keV}^{B_{LY}}
\end{eqnarray}
with $a_{LY}=9/5$ and $B_{LY}=4/5$. This relation was fit to the
observed \Yx\ and \Lx\ values for the sample, with $B_{LY}$ and
$C_{LY}$ as free parameters and the luminosities divided by
$E(z)^{9/5}$. The best fitting relation was measured with an orthogonal,
weighted ``BCES'' regression \citep[as described by][]{akr96}, on the data
in log space. The intrinsic scatter was then measured in the following
way. Consider a set of data ($x_i$,$y_i$) with
measurement uncertainties ($\sigma(x)_i$,$\sigma(y)_i$) fit by a straight
line model of slope $m$ and intercept $c$. The intrinsic
scatter of the data in the y direction ($\sigma_y$), is defined as the value
of $\sigma_y$ that gives a reduced \chisq\ statistic of unity. The value of
\chisq\ is given by
\begin{eqnarray}\label{e.chi}
\chisq & = & \sum_{i}\frac{(y_i-c-mx_i)^2}{\sigma^2(y)_i + m^2\sigma^2(x)_i
+ \sigma^2_y}.
\end{eqnarray}

With this method, the intrinsic scatter in luminosity, $\sigma_L$ was
measured for each relation. As the data were fit in log space, $\sigma_L$
was multiplied by the natural log of 10 to give the scatter as a fraction
of \Lx. This method was also used to measure the intrinsic scatter in the
x variable (\Yx, kT or \Mf). The uncertainty on the measured sample was
estimated from bootstrap resamples of the data.

The \LY\ relations obtained with luminosities measured in different
apertures are summarised in Table \ref{t.yl}. The apertures used correspond
to three levels of cooling core correction, but note that the
$(0.15<r<1)\rf$ aperture was used to measure the temperature for \Yx\ in
all cases. The full $r<\rf$ aperture includes all emission from cool cores,
and the $70\kpc<r<\rf$ aperture is commonly used to exclude the strongest
parts of cooling cores, with the resulting luminosity scaled by $1.06$ to
account for the excluded non-cool-core emission
\citep[\egc][]{mar98a}. Finally, the $(0.15<r<1)\rf$ aperture was chosen to
conservatively excluded all cooling core emission. Bolometric luminosities
were used for these relations.

\begin{figure}
\rotatebox{-90}{\scalebox{0.33}{\includegraphics{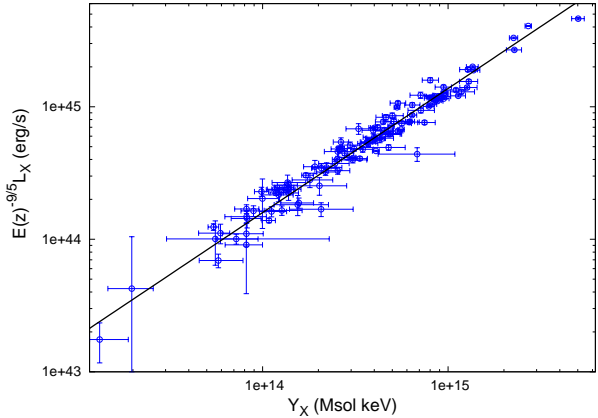}}}
\caption{\label{f.yl}{\LY\ relation for bolometric luminosities measured in
the $(0.15<r<1)\rf$ aperture. Luminosities are scaled by the predicted
self-similar evolution.}}
\end{figure}

As Table \ref{t.yl} demonstrates, the scatter in the \LY\ relation is
significantly reduced by excluding cool cores, and the $(0.15<r<1)\rf$
aperture is the most effective at reducing scatter. The relation derived
using this aperture is plotted in Fig. \ref{f.yl}. 

\begin{deluxetable*}{p{1cm}lllllll}
\setlength{\tabcolsep}{0.75\tabcolsep}
\tabletypesize{\scriptsize}

\tablecaption{\label{t.yl}Summary of the best fitting scaling relations
between luminosity and \Yx, kT, and \Mf\ (the X variables listed in
column 1). Luminosities were measured in different apertures, either from
spectral fits or from count rates as discussed in the text and listed in
columns 2 and 3. The energy bands used for the soft band luminosities are given in the clusters' rest frames. Relations of the form $\Lx=CE(z)^\alpha(X/X_{*})^B$ were
fit to the data with $X_*=(6\keV, 4\times10^{14}\Msol\keV,
4\times10^{14}\Msol)$ and $\alpha=(9/5, 1, 7/3)$ (self-similar evolution)
for \Yx, kT, and \Mf\ respectively. Column 4 gives the redshift range of the clusters included in the fit. Columns 5 and 6 give the normalisation
and slope of the $\mbox{\ensuremath{\mathrm{L_{X}-X}}}$ relation in
question, and columns 7 and 8 give the intrinsic fractional scatter in \Lx\
and X. Masses were estimated from \Yx\ and the parameters of the
best-fitting \LM\ relations include uncertainties and scatter in the \YM\
relation.}

\tablehead{
\colhead{X} & \colhead{\Lx\ type} & \colhead{\Lx\ aperture} & \colhead{redshift} & \colhead{C} & \colhead{$B$} & \colhead{$\sigma_L$} & \colhead{$\sigma_X$}\\
\colhead{} & \colhead{} & \colhead{} & \colhead{} & \colhead{$(10^{44}\ergps)$} & \colhead{} & \colhead{} & \colhead{}
}

\startdata
\Yx\	   & spectral, bolometric 			& $r<\rf$	 		& $0.1<z<1.3$	& $10.0\pm0.4$		& $1.10\pm0.04$ & 0.36$\pm0.03$	& 0.33$\pm0.03$\\ 
\Yx\	   & spectral, bolometric 			& $70\kpc<r<\rf$  		& $0.1<z<1.3$	& $8.7\pm0.2$		& $1.06\pm0.03$ & 0.21$\pm0.02$	& 0.20$\pm0.01$\\
\Yx\	   & spectral, bolometric 			& $(0.15<r<1)\rf$ 		& $0.1<z<1.3$	& $5.8\pm0.1$		& $0.94\pm0.03$ & 0.11$\pm0.02$	& 0.12$\pm0.02$\\
\Yx\	   & spectral, bolometric 			& $(0.15<r<1)E(z)^{-2/3}\Mpc$ 	& $0.1<z<1.3$	& $7.0\pm0.1$ 		& $1.00\pm0.02$ & 0.13$\pm0.02$	& 0.14$\pm0.02$\\
\Yx\	   & spectral, $(0.5-2)\keV$ 	& $(0.15<r<1)E(z)^{-2/3}\Mpc$ 	& $0.1<z<1.3$	& $1.85\pm0.04$ 	& $0.84\pm0.03$ & 0.20$\pm0.01$	& 0.24$\pm0.03$\\
\Yx\	   & count rate, $(0.5-2)\keV$ 	& $(0.15<r<1)E(z)^{-2/3}\Mpc$ 	& $0.1<z<1.3$	& $1.91\pm0.04$ 	& $0.82\pm0.03$ & 0.19$\pm0.01$	& 0.23$\pm0.02$\\ \hline
kT	   & spectral, bolometric 			& $(0.15<r<1)\rf$ 		& $0.1<z<1.3$	& $6.6\pm0.3$ 		& $2.8\pm0.2$ 	& 0.34$\pm0.04$	& 0.12$\pm0.02$\\ \hline
\Mf\	   & spectral, bolometric 			& $r<\rf$ 			& $0.1<z<1.3$	& $5.6\pm0.3$		& $1.96\pm0.10$ & 0.39$\pm0.04$	& 0.21$\pm0.01$\\
\Mf\	   & spectral, bolometric 			& $(0.15<r<1)\rf$ 		& $0.1<z<1.3$	& $3.5\pm0.1$		& $1.63\pm0.08$ & 0.17$\pm0.02$	& 0.08$\pm0.02$\\
\Mf\	   & count rate, $(0.5-2)\keV$ &	$(0.15<r<1)E(z)^{-2/3}\Mpc$ 	& $0.1<z<1.3$	& $1.20\pm0.05$ 	& $1.45\pm0.07$ & 0.21$\pm0.02$	& 0.16$\pm0.02$\\
\Mf\	   & count rate, $(0.5-2)\keV$ &	$r<1E(z)^{-2/3}\Mpc$ 		& $0.1<z<0.5$	& $1.7\pm0.1$ 		& $1.7\pm0.1$ 	& 0.46$\pm0.05$	& 0.28$\pm0.02$\\
\Mf\	   & count rate, $(0.5-2)\keV$ &	$r<1E(z)^{-2/3}\Mpc$ 		& $0.5<z<1.3$	& $1.6\pm0.1$ 		& $1.5\pm0.1$ 	& 0.19$\pm0.05$	& 0.16$\pm0.04$\\
\enddata

\end{deluxetable*} 

\subsection{The effect of cooling cores and substructure}\label{sec:effect-cooling-cores}
\begin{figure}
\rotatebox{-90}{\scalebox{0.33}{\includegraphics{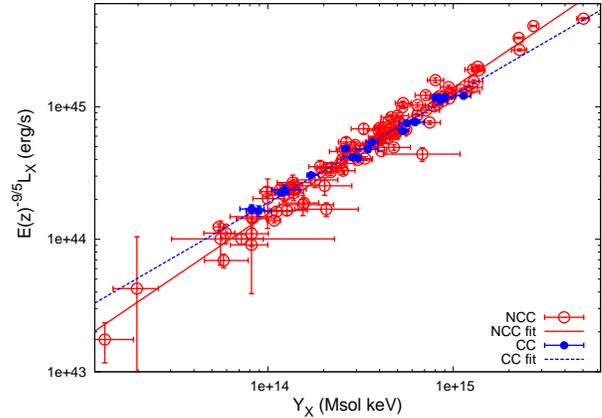}}}
\caption{\label{f.cf}{\LY\ relations for cool core (CC) and non-cool core
(NCC) clusters. Luminosities are bolometric, measured in the
$(0.15<r<1)\rf$ aperture, and scaled by the predicted self-similar
evolution.}}
\end{figure}

\begin{figure}
\rotatebox{-90}{\scalebox{0.33}{\includegraphics{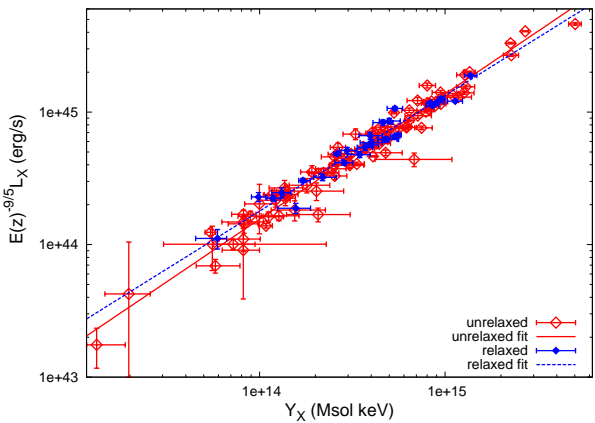}}}
\caption{\label{f.morpho}{\LY\ relations for relaxed and unrelaxed
clusters. Luminosities are bolometric, measured in the 
$(0.15<r<1)\rf$ aperture, and scaled by the predicted self-similar
evolution.}}
\end{figure}

In order to test for any residual effects of cooling core emission on the
\LY\ relation, the sample was split into cool core (CC) and non-cool core
(NCC) subsamples. Clusters were classed as CC if the temperature measured
within $r<0.15\rf$ was cooler than that measured within $(0.15<r<0.3)\rf$
at a significance of $>2\sigma$. While this method is dependent on the
quality of the data used and so is not a robust classification scheme, it
allows us to separate out the strongest cool core clusters to look for
departures from the population. The relation was fit to the two subsets
separately, and the relations are plotted in Fig. \ref{f.cf}. There is no
evidence for any offset between the CC and NCC populations, though the slope of
the CC relation is $\sim3\sigma$ shallower than the NCC relation. The
intrinsic scatter of the 18 CC clusters was $\sigma_L=0.04\pm0.02$, significantly
lower than that of the NCC clusters ($\sigma_L=0.13\pm0.02$). This is likely a
reflection of the fact that CC clusters are generally more relaxed than the
rest of the population. When core emission was not excluded, the CC
clusters were significantly offset to higher \Lx, as expected.
The sample contains clusters in a wide variety of dynamical states from
relaxed, cool core clusters to early and late stage mergers with a range of
mass ratios \citep{mau07a}. To further investigate the effect of cluster
morphology on the \LY\ relation, the clusters were split into relaxed and
unrelaxed subsamples based on their measured centroid shift ($\langle w
\rangle$) parameters \citep[see][for details]{mau07a}. $\langle w
\rangle$ was calibrated by visual inspection of cluster images, and the 28
clusters with $\langle w \rangle<0.005\rf$ were classed as relaxed, with
the remaining 87 classed as unrelaxed. The \LY\ relation was fit to these
subsets separately and the data are plotted in Fig. \ref{f.morpho}. The
normalisations were consistent for both subsets, and the slope was slightly
shallower ($\sim1\sigma$) for the relaxed clusters, a similar trend to the
CC and NCC clusters. The scatter was slightly, but not significantly, lower
for the relaxed clusters. We note that the CC clusters generally have the
lowest $\langle w \rangle$, and represent the most relaxed subset of the
sample.

\subsection{Evolution}\label{s.evol}
For the previous fits, we assumed that the \LY\ relation evolves
self-similarly with $a_{LY}=9/5$. We have tested the validity of this
assumption by using the data to determine the best-fitting value of
$a_{LY}$. For different values of $a_{LY}$, luminosities were divided by
$E(z)^{a_{LY}}$, the best fitting relation was found via BCES
regression and the \chisq\ was computed using equation \ref{e.chi} (with
$\sigma_L=0$). The best-fitting $a_{LY}$ was the value that minimised
\chisq, and the $1\sigma$ confidence intervals were given by the values
that gave $\Delta\chisq=1$ from the minimum.

\begin{figure}
\rotatebox{-90}{\scalebox{0.33}{\includegraphics{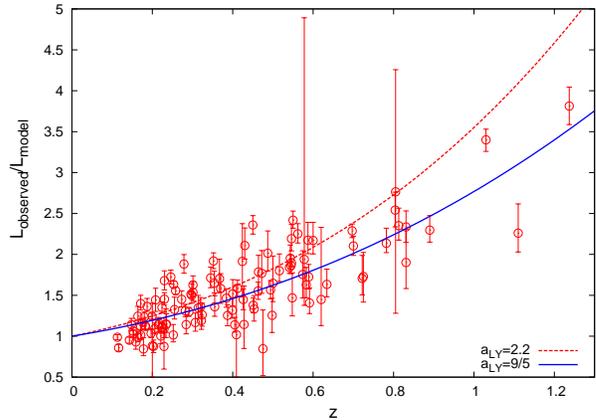}}}
\caption{\label{f.evol}{Ratio of the observed $L_X$ to that predicted by
the local \LY\ relation is plotted against redshift. The curves show the
predicted self-similar (solid) and best-fitting (dashed) evolution.}}
\end{figure}

The best-fitting evolution was $a_{LY}=2.2\pm0.1$, significantly
($4\sigma$) higher than the self-similar value. To illustrate the
evolution, a local \LY\ relation was defined by fitting the 20 clusters
at $0.1<z<0.2$. Self-similar evolution was assumed for measuring this local
relation, though the effect at these low redshifts is small. The ratio of a
cluster's measured luminosity to that predicted by this local relation for
the cluster's $Y_X$ is then equal to $E(z)^{a_{LY}}$. Fig. \ref{f.evol}
shows the value of this ratio for each cluster along with the loci of the
self-similar and best-fitting evolution.

While the data prefer a stronger evolution than the self-similar
prediction, both evolution models give an unacceptable fit
($\chisq/\nu\approx4$). This is a result of the significant intrinsic
scatter in the data; both evolution models require $\sigma_L=0.11$
to give a reduced $\chisq$ of unity. Without a priori knowledge of the
details of this scatter that would allow us to distinguish between
evolution models, we find no compelling reason to abandon the
theoretically motivated self-similar evolution, and continue to adopt it
throughout this work. 

\subsection{Survey-quality data}\label{s.surv}
The tight correlation between \Lx\ and \Yx\ and that between \Yx\ and mass
(KVN06) suggests that \Lx\ can provide an effective mass proxy. However,
data that are of sufficient quality ($\sim1000$ net counts) to allow a
spectrum to be fit and a luminosity measured in that way, will generally
allow \Yx\ to be measured directly (as is the case for our sample). The
prospect of estimating masses reliably from luminosities has much greater
potential for clusters detected in serendipitous cluster surveys, where
only count rates are available. We now investigate how well the \LY\
relation holds up for such relatively low-quality data.

Up to this point, \Yx\ has weakly influenced our \Lx\ measurements, as \rf\
(and hence the luminosity aperture) was estimated from \Yx. This dependence
was removed by switching to an aperture of $(0.15<r<1)E(z)^{-2/3}\Mpc$,
chosen to approximate $(0.15<r<1)\rf$. Encouragingly, the scatter in the
\LY\ relation was insensitive to this change, with $\sigma_L=0.13\pm0.02$. Next,
the energy band of the luminosities measured from the spectral fits was
changed from the bolometric band to the $(0.5-2)\keV$ band in each
cluster's rest frame. This change was motivated by the fact that the soft
band \Lx\ is less sensitive to the cluster temperature than the bolometric
band, so can be more reliably estimated from a soft band count rate
(typical of survey data) without a measured temperature. The soft band
\LY\ relation showed slightly increased scatter ($\sigma_L=0.21\pm0.02$) likely
due to the fact that the increased temperature dependence of the bolometric
\Lx\ suppresses the scatter somewhat, since \Yx\ is proportional to
temperature. 

\begin{figure}
\rotatebox{-90}{\scalebox{0.40}{\includegraphics{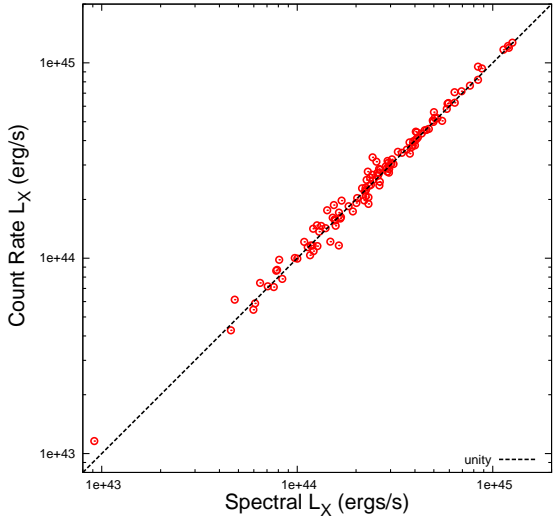}}}
\caption{\label{f.lxlx}{The rest-frame $(0.5-2.0)\keV$ luminosity measured
from a full spectral fit is plotted against the luminosity estimated from
the count rate in the same energy band. The $(0.15<r<1)E(z)^{-2/3}\Mpc$
aperture was used for both measurements and the dashed line indicates a
one-to-one correspondence.}}
\end{figure}

Finally, for each cluster we measured the net count rate in the observed
frame $(0.5-2)\keV$ band, and converted it to a rest frame $(0.5-2)\keV$
luminosity. For the conversion we assumed an absorbed APEC spectral model with the
absorption fixed at the galactic value for each cluster, the redshift fixed
at that of the cluster, and metal abundance fixed at $0.3\Zsol$. The
temperature of the spectral model was set at an initial guess and \Lx\
was calculated. A luminosity-temperature relation was then used to estimate
$kT$ from \Lx\ and the process was iterated until \Lx\ stabilised. The
\citet{mar98a} \LT\ relation was used, with self-similar evolution, but the
conversion is insensitive to the choice of \LT\ relation. The reliability
of this conversion was verified by comparing our estimated luminosities
with those measured from the spectral fits, and an excellent agreement was
found (illustrated in Fig. \ref{f.lxlx}). The \LY\ relation was then
measured using these estimated soft band luminosities, and the scatter did
not change significantly $\sigma_L=0.19\pm0.01$.

\section{The \YM\ relation}
The ultimate goal of this study is to investigate the \LMf\ relation using
\Yx\ as a mass proxy. The use of \Yx\ as a low-scatter mass proxy has
strong support from simulations \citep[KVN06][]{poo07}, but there have not
yet been many observational studies of this parameter. The \YM\ relation
measured for the V06 clusters exhibits the same low scatter found in the
simulated relation (KVN06), but the predicted self-similar evolution of the
relation has not yet been tested observationally. It is therefore
desirable to investigate the scatter and evolution of the \YM\ relation
using clusters from our sample. This requires independent mass estimates
for for the clusters. These can be estimated from temperature and gas
density profiles measured from X-ray observations, assuming hydrostatic
equilibrium. However, data of sufficient quality are unavailable for the
majority of our sample, and a full mass analysis of the remainder is beyond
the scope of the current work. Instead, we searched the literature for mass
estimates for clusters in our sample. We found 12 clusters in our sample
with values of \Mf\ estimated from full X-ray hydrostatic mass analyses
(i.e. using temperature and density profiles rather than an isothermal
approximation). Masses for 3 clusters were taken from the overlap between
our sample and the V06 sample, one was obtained from the \XMM\
sample of \citet{arn05}, and the remainder came from the \XMM\ and
\Chandra\ samples presented by \citet{kot05} and \citet{kot06}
respectively. The properties of the clusters are summarised in Table
\ref{t.yml}.

\begin{deluxetable*}{llllllll}
\setlength{\tabcolsep}{0.75\tabcolsep}
\tabletypesize{\scriptsize}

\tablecaption{\label{t.yml}Properties of the subset of clusters with masses
based on high quality data available in the literature. Masses were
estimated using X-ray hydrostatic mass analyses within \rf. The luminosities
and \Yx\ values presented were measured from our analyses, with \Lx\
measured in the  $(0.15<r<1)\rf$ aperture. The references
for the mass estimates listed in column 6 correspond to: 1 - \citet{vik06a},
2 - \citet{arn05}, 3 - \citet{kot05}, 4 - \citet{kot06}.}

\tablehead{
\colhead{Cluster} & \colhead{z} & \colhead{\Lx} & \colhead{\Yx} & \colhead{\Mf} & \colhead{Reference} \\
\colhead{} & \colhead{} & \colhead{$(10^{44}\ergps)$} & \colhead{$(10^{14}\Msol\keV)$} & \colhead{$(10^{14}\Msol)$} & \colhead{}
}

\startdata
A1413	& 0.143	& $7.6\pm0.1$	& $5.6\pm0.2$	& $7.8\pm0.8$	& 1\\
A907	& 0.153	& $4.7\pm0.1$	& $3.0\pm0.1$	& $4.7\pm0.4$	& 1\\
A383	& 0.187	& $3.6\pm0.1$	& $1.7\pm0.1$	& $3.1\pm0.3$	& 1\\
A2204	& 0.152	& $12.7\pm0.3$	& $8.9\pm0.7$	& $8.4\pm0.8$	& 2\\
MS0302.7+1658	& 0.424	& $3.0\pm1.2$	& $9.2\pm1.5$	& $2.2\pm0.6$	& 3\\
MS0015.9+1609	& 0.541	& $33.9\pm0.6$	& $13.4\pm1.0$	& $8.8\pm1.1$	& 3\\
CLJ1120+4318	& 0.600	& $8.3\pm0.7$	& $2.5\pm0.6$	& $4.6\pm1.1$	& 3\\
MACSJ0159.8-0849& 0.405	& $17.7\pm0.3$	& $11.2\pm1.1$	& $9.9\pm3.1$	& 4\\
MACSJ0329.6-0211& 0.450	& $11.3\pm0.6$	& $3.3\pm0.4$	& $3.6\pm1.0$	& 4\\
RXJ1347.5-1145	& 0.451	& $41.4\pm0.8$	& $2.4\pm0.2$	& $13.8\pm2.8$	& 4\\
MACSJ1621.3+3810& 0.463	& $8.5\pm0.5$	& $3.8\pm0.4$	& $4.4\pm1.0$	& 4\\
MACSJ1423.8+2404& 0.543	& $11.3\pm0.7$3	& $4.1\pm0.7$	& $4.6\pm1.0$	& 4\\

\enddata

\end{deluxetable*} 

\begin{figure}
\rotatebox{-90}{\scalebox{0.33}{\includegraphics{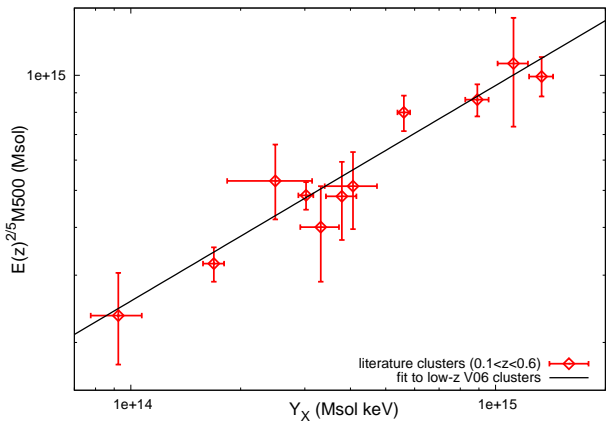}}}
\caption{\label{f.litYM}{\YM\ relation using our measured \Yx\ values with
masses taken from the literature. The masses are scaled by the expected
self-similar evolution. The line shows the best fit to the V06 data; this
line is not fit to the data plotted here, but 3 of the datapoints are in
common with the V06 sample.}}
\end{figure}

The \YM\ relation for these 12 clusters is shown in Fig. \ref{f.litYM}. It
was constructed using our measured \Yx\ values, with the masses taken from
the literature and scaled by the predicted self-similar evolution
($E(z)^{2/5}$). These data were compared with the best fit relation to the
V06 clusters (equation \ref{e.ym}), with which three clusters are in
common. The V06 relation is a good description of the data; the reduced
\chisq\ is $<1$ so there is no measurable intrinsic scatter. While the
intrinsic scatter cannot be well constrained with this small sample, the
tight correlation in Fig. \ref{f.litYM} adds further support to the use of
\Yx\ as a low-scatter mass proxy, and provides the first observational
support for the self-similar evolution of the \YM\ relation to $z=0.6$.

\section{Estimating masses from luminosities}\label{sec:estim-mass-from}
As a final exercise, we use the \YM\ relation to estimate masses for the
clusters in the sample, and then investigate the \LMf\ relation using
luminosities measured in different ways.  In order to include the effects
of both the intrinsic scatter and the uncertainties on the \YM\ relation on
our derived \LMf\ relation, we used a Monte-Carlo approach as follows. A
random realisation of the \YM\ relation (equation
\ref{e.ym}) was created by drawing values of the slope and normalisation
from the Gaussian distributions defined by $B_{YM}=0.564\pm0.009$ and
$C_{YM}=(7.047\pm0.097)\times10^{14}\Msol$. These are the best-fitting
values from the V06 data, but with the errors taken from the fractional
errors on the KVN06 simulated relation. For each cluster, a mass was
computed using this relation (assuming self-similar evolution) and that
mass was then randomised under a Gaussian in natural log space centred at
the true value with $\sigma=0.071$. This corresponds to the $7.1\%$
intrinsic scatter in mass in the \YM\ relation found by KVN06.

\begin{figure}
\rotatebox{-90}{\scalebox{0.33}{\includegraphics{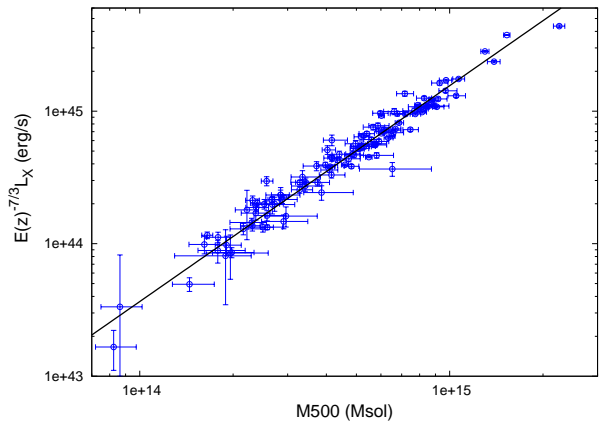}}}
\caption{\label{f.lm}{\LMf\ relation for the clusters with masses estimated
from the \YM\ relation and bolometric luminosities measured from spectral
fits in the $(0.15<r<1)\rf$ aperture, and scaled by the predicted
self-similar evolution.}}
\end{figure}

The \LMf\ relation was then fit using these randomised masses and our
measured luminosities. The process was repeated 1000 times and the
distributions of measured slope ($B_{LM}$), normalisation ($C_{LM}$) and
scatter in \LMf\ were then used to determine the values and uncertainties
of those parameters. The uncertainties and scatter in the \YM\ relation
did not contribute strongly to those on the derived \LMf\ relations. In all
cases, the $\pm34$ percentiles about the mean of the $B_{LM}$ and $C_{LM}$
distributions derived from the Monte-Carlo randomisations enclosed a
smaller range than the statistical uncertainties from the regression fit to
the unperturbed \LMf\ relation. The Monte-Carlo uncertainties were added in
quadrature to the regression uncertainties to give the final uncertainties
for these parameters. We used the median and $\pm34$ percentiles of the
distribution of $\sigma_L$ from the Monte-Carlo runs as our estimate of the
scatter in the \LMf\ relation. This too was not strongly affected by the
scatter in \YM; the median $\sigma_L$ was $\lta0.03$ higher than that
measured for the unperturbed \LMf\ data in all cases. The measured \LMf\
relations are summarised in Table \ref{t.yl} and the relation obtained for
bolometric \Lx\ measured from spectral fits in the $(0.15<r<1)\rf$ aperture
is plotted in Fig. \ref{f.lm}. As expected, the scatter follows the same
trend as in the \LY\ relation, reducing from $0.39\pm0.04$ to $0.17\pm0.02$
when the core regions were fully excluded, with a slightly larger scatter of
$\sigma_L=0.21\pm0.02$ for survey-quality data.

\subsection{Correlation of scatter in the \LY\ and \YM\ relations.}\label{sec:corr-scatt-ly}
\begin{figure}
\rotatebox{-90}{\scalebox{0.40}{\includegraphics{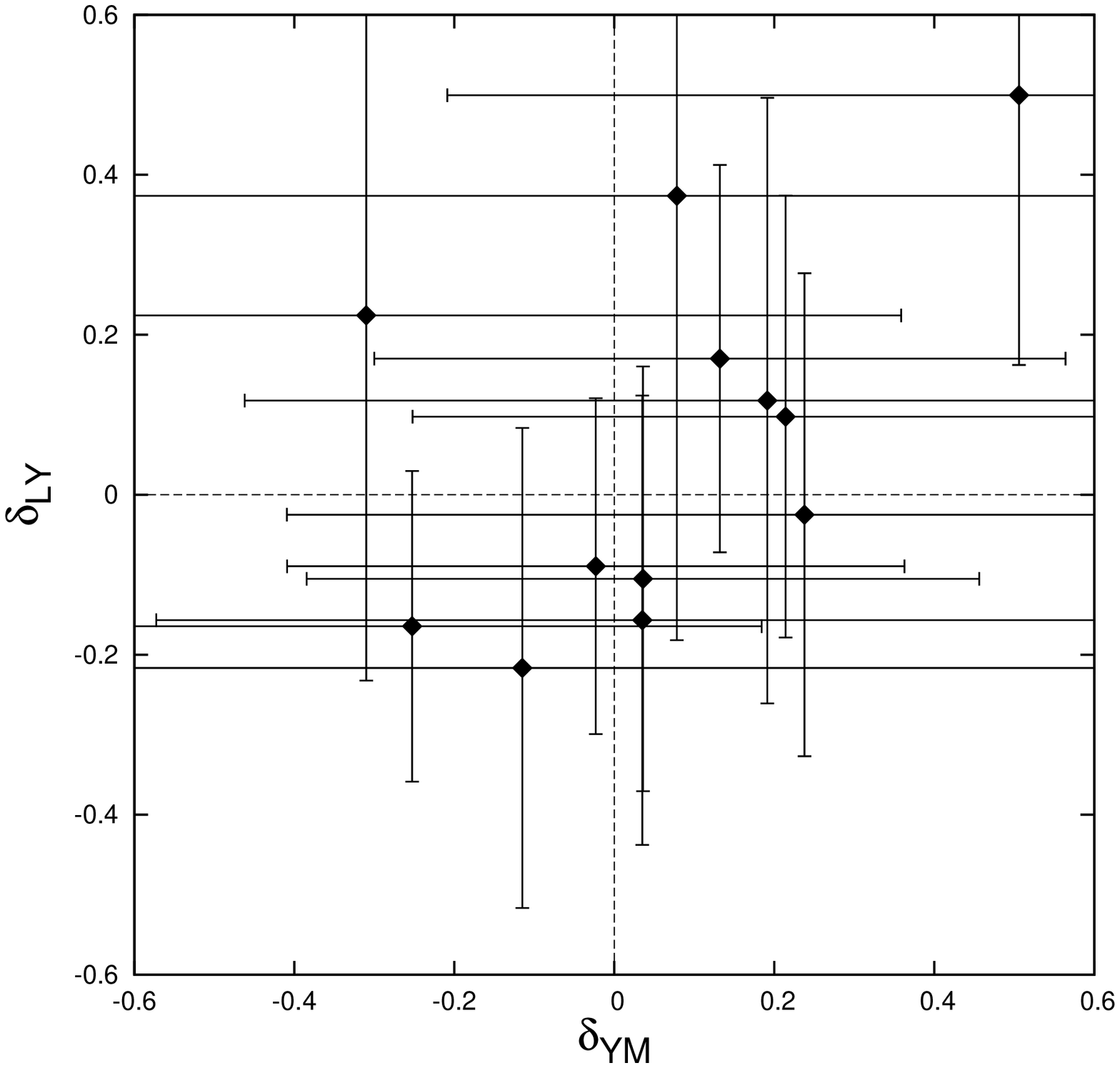}}}\\
\rotatebox{-90}{\scalebox{0.40}{\includegraphics{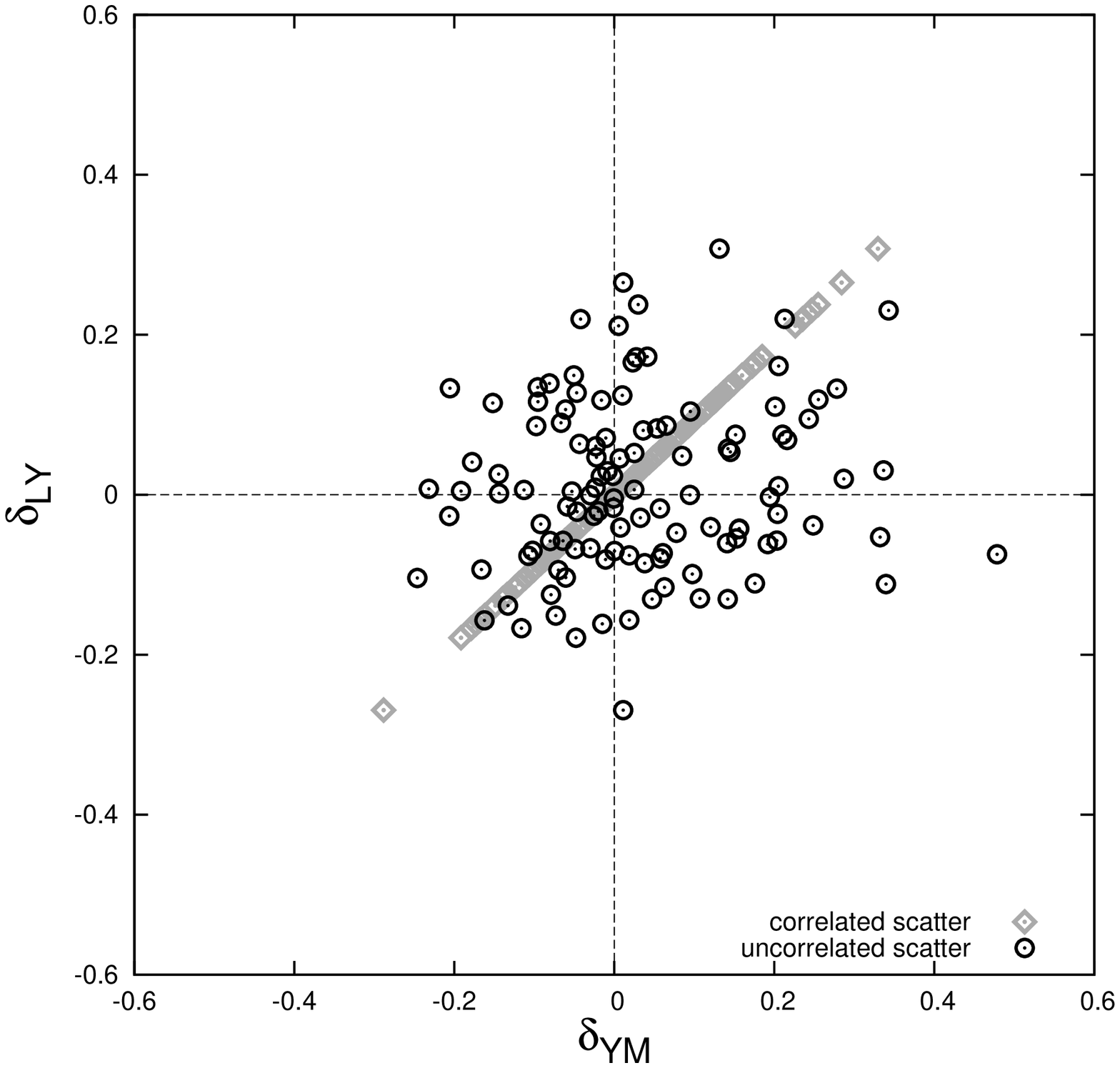}}}
\caption{\label{f.dydl}{{\it Left:} Fractional residuals in \Lx\ from
the \LY\ relation ($\delta_{LY}$) and in \Yx\ from 
the \YM\ relation ($\delta_{YM}$) for the subset of clusters with mass
estimates taken from the literature. {\it Right:} The same residuals for
the full sample from one of the Monte-Carlo runs for the assumptions of
correlated and uncorrelated scatter in the \LY\ and \YM\ relations.}}
\end{figure}

Our Monte-Carlo method assumes that the scatter in the \LY\ and \YM\
relations is uncorrelated. If they were positively or negatively correlated,
then the derived scatter on the \LMf\ relation would be under- or
overestimated respectively. Ideally, one would measure the offset of
clusters in \Lx\ from the \LY\ relation ($\delta_{LY}$) and in \Yx\ from
the \YM\ relation ($\delta_{YM}$) and test for correlations between
those offsets. This is not possible for the full sample, as independent
mass estimates are not available, but this can be investigated for the
subset of clusters with masses taken from the literature. The offsets for
these clusters are plotted in Fig. \ref{f.dydl}. The data hint at a
positive correlation, but the correlation is not significant if the
uncertainties on the data points are included.

In order to investigate the case of positive correlation between the
scatter in the \LY\ and \YM\ relations and thus place a conservative upper
limit on the \LMf\ scatter, we rederived the \LMf\ relation assuming that
$\delta_{LY}$ and $\delta_{YM}$ are perfectly correlated. In this process,
cluster masses were estimated using the \YM\ relation as before, but were
then scattered to give a $\delta_{YM}$ that was proportional to the
cluster's known $\delta_{LY}$. The constant of proportionality used was the
ratio of the scatter in the \LY\ and \YM\ relations. This is illustrated in
Fig. \ref{f.dydl} along with an example of uncorrelated scatter as assumed
previously. With this assumption of correlated scatter, the resulting
scatter in the \LMf\ relation for \Lx\ measured from spectral fits in the
$(0.15<r<1)\rf$ aperture increased from $\sigma_L=0.17\pm0.02$ to
$\sigma_L=0.23\pm0.02$. For comparison, the \LMf\ relation was also
constructed for the subset of 12 clusters with independent mass estimates
from the literature, using our measured luminosities in the $(0.15<r<1)\rf$
aperture. The scatter of these data about the \LMf\ relation derived for
the full sample was found to be $\sigma_L=0.19\pm0.11$. The large
uncertainties on the scatter due to the small number of clusters with
independent mass measurements limit our ability to pin down the extent of
the correlation between $\delta_{LY}$ and $\delta_{YM}$, and the precise
value of the scatter in the \LMf\ relation. This will be addressed in
future work by deriving masses for a larger subset of our sample. In the
mean time, we continue with the assumption of uncorrelated scatter, but the
increase in $\sigma_L$ for correlated scatter indicates the size of any
sytematic effect due to this assumption.

\subsection{The effect of including core emission for high-z clusters}\label{sec:effect-incl-core}
A source of concern for the application of this method is that a
substantial fraction of the cluster emission is excluded by excluding the
cluster core out to $0.15\rf$ (or $0.15E(z)^{-2/3}\Mpc$). To quantify this,
we compared the flux measured from spectral fits in the $(0.15<r<1)\rf$
aperture and the total flux within \rf\ after correction for cool
cores. The standard cool-core correction of multiplying the flux measured
in the $70\kpc<r<\rf$ aperture by $1.06$ was used \citep{mar98a}. The mean
ratio of the $(0.15<r<1)\rf$ to total flux was $0.70$, so typically $30\%$
of the flux is excluded. A second issue is that for telescopes with a
significant point spread function (PSF) such as \ROSAT\ and \XMM, the
central exclusion region becomes smaller than the PSF for distant
clusters, making core exclusion difficult. 

Recently \citep{vik06c} used measurements of the gas density profile slopes
in the cores of galaxy clusters to show that the fraction of clusters with
cool cores is low at $z\ga0.5$. As cool cores are the dominant source of
scatter in these scaling relations, it is therefore interesting to measure the
scatter without core exclusion for distant clusters to determine if
it can be neglected for high-z clusters. Luminosities were
estimated from count rates in the $r<1E(z)^{-2/3}\Mpc$ aperture (\ie
including core emission) and the sample was split at $z=0.5$ into low- and
high-redshift subsets. The \LMf\ relation was calculated for each subset as
described above, and the parameters are given in Table \ref{t.yl}. The
best-fitting relations are consistent for the two subsets, but the scatter
for the low redshift subset ($\sigma_L=0.46\pm0.05$) is significantly
larger than that for the high-z clusters ($\sigma_L=0.19\pm0.05$). In fact,
the scatter for the distant clusters with the cores included is the same as
that for the full sample with cores excluded. This suggests that the core
exclusion is generally not required for clusters at $z>0.5$. Note that
while the measurement errors are generally larger for the more distant
clusters, these values for the {\it intrinsic} scatter take those into
account.

\section{Discussion}
Table \ref{t.yl} shows that the scatter in the scaling relations is
dominated by the cluster core regions. The obvious candidate is the
enhanced \Lx\ due to cool cores. However, it should be noted that the
strongest effects of mergers on \Lx\ occur around core passage of the
merging bodies \citep[e.g.][]{ran02}. By excluding a large core
region, we also effectively remove the strongest part of the scatter due to
mergers. The scatter from the core regions is thus due to a combination of
mergers and cooling. Once the core regions are excised, the morphological
status of the clusters has little effect on the scatter unless only the
most relaxed systems (the CC clusters) are considered (\textsection
\ref{sec:effect-cooling-cores}). The total intrinsic scatter of
$\sigma_L=0.36$ in the \LY\ relation can thus be resolved into different
components. The total scatter reduces by $\approx70\%$ when the central
$0.15\rf$ is excluded. A further $\approx20\%$ of the scatter is removed
when only the most relaxed CC systems are considered; this portion of the
scatter is likely due to merger effects outside the cores. The remaining
$\approx10\%$ of the scatter is likely due to residual substructure even in
the most relaxed systems, to the scatter in the \YM\ relation, which has a
weak effect on our measured luminosities through \rf, and to contributions
from unresolved point sources.

For comparison with the \LY\ relation, we also measured the \LT\ relation
for the sample, using luminosities and temperatures measured in the
$(0.15<r<1)\rf$ aperture. The best fit parameters are given in Table
\ref{t.yl}, but the main point to note is that even with our most
conservative core exclusion, $\sigma_L$ in the \LT\ relation is three times
that in the corresponding \LY\ relation. The fact that \Yx\ has been shown
to be a low-scatter mass proxy implies that the larger scatter in the
\LT\ relation is due to scatter in the \MT\ relation. As our mass estimates
are based on kT (via \Yx), we cannot investigate the scatter in the \MT\
relation for our sample. However, our results suggest that for samples
such as ours, which include relaxed and disturbed clusters, luminosity
(with a large core region excluded) has a tighter relation with mass than
temperature does. Although, recall that \Yx\ has a tighter relation with
mass than do either \Lx\ or kT. Clusters with strong deviations from the
\LT\ relation will be studied in detail in a future paper.

\subsection{The shape of the scatter in the \LY\ relation}\label{sec:form-scatter-ly} 
\begin{figure}
\rotatebox{-90}{\scalebox{0.33}{\includegraphics{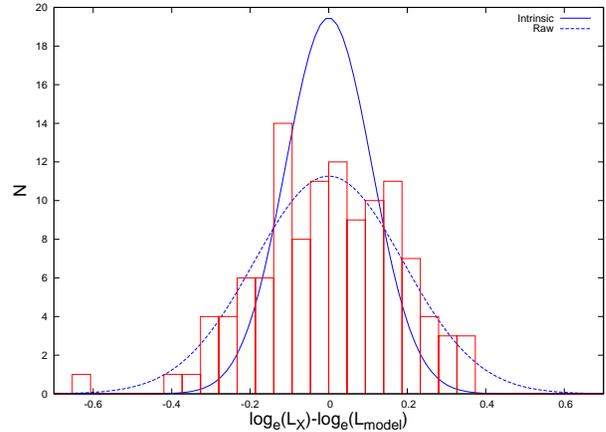}}}
\caption{\label{f.scat}{Histogram of the residuals from the \LY\ relation
in $\log_{e}$ space. This raw scatter is well described by a Gaussian of
$\sigma_L=0.19$ (dashed line). The solid line is a Gaussian with
$\sigma_L=0.11$ corresponding to the measured intrinsic scatter.}}
\end{figure}

It is desirable to adopt a functional form for the intrinsic scatter in
the scaling relations in order to include its effects on \eg sample
completeness and derived mass functions. With 115 clusters it is possible
to investigate the form of the scatter in the \LY\
relation. Fig. \ref{f.scat} shows a histogram of the $\log_{e}$ space
residuals from the \LY\ relation fit to all clusters, using spectral \Lx\
measured in the $(0.15<r<1)\rf$ aperture. This raw scatter is well
described by a Gaussian in log space with $\sigma_L=0.19$; a
Kolmogorov-Smirnov test gives a null hypothesis probability of $0.99$ that
the data are consistent with that Gaussian. Here we have implicitly
assumed that the scatter does not vary with redshift.

\begin{figure}
\rotatebox{-90}{\scalebox{0.33}{\includegraphics{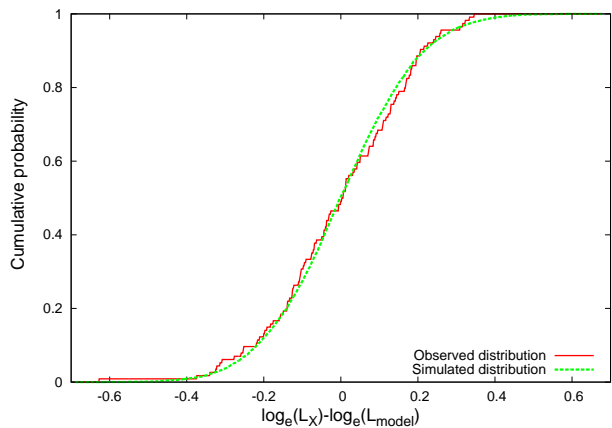}}}
\caption{\label{f.ks}{Cumulative probability distributions for the observed
(solid line) and simulated (dashed line) raw scatter.}}
\end{figure}

The form of the intrinsic scatter is harder to recover. For each cluster we
have an observed luminosity (\Lx) and a luminosity predicted by the
\LY\ relation for that clusters measured \Yx\ ($L_{model}$). Each log space
residual ($\log_{e}(\Lx/L_{model})$) is associated with a measurement
error on ($\Lx/L_{model}$), which we assume is described by a Gaussian
distribution in linear space. The raw scatter can then be approximated as a
convolution of the intrinsic scatter distribution with a linear Gaussian
representing the mean measurement error on the residuals. This
approximation was used to test if the intrinsic scatter could be described
by a lognormal distribution. A simulated residual point was drawn from
under a Gaussian with $\sigma=0.11$ in log space (our intrinsic
scatter model). This value of $\log_{e}(\Lx/L_{model})$ was transformed
into linear space ($\Lx/L_{model}$) and then randomised under a
Gaussian of $\sigma=0.17$ (the mean measurement error on
$\Lx/L_{model}$). Note that we verified that the measurement errors on
$\Lx/L_{model}$ were not significantly correlated with
$\Lx/L_{model}$. This was repeated for $10,000$ simulated residuals to
give a simulated raw scatter distribution. The resulting distribution was
compared to the observed raw scatter distribution using the
Kolmogorov-Smirnov test, and the cumulative probability distributions are
plotted in Fig. \ref{f.ks}. The distributions agree very well, with a null
hypothesis probability of $0.91$, indicating that the intrinsic scatter is
well described by a lognormal distribution.

\subsection{The effect of Eddington bias}\label{sec:effect-malquist-bias}
An important factor that we have not yet addressed is the issue of
Eddington bias. In a flux-limited sample, clusters of a range of masses
that have fluxes near the detection limit are scattered into and out of the
sample, due to the intrinsic scatter in cluster luminosities for a given
mass. Because the mass function is a decreasing function of mass, more
clusters will scatter into than out of the sample. This results in a bias
in both the number of clusters and their mean luminosities (as clusters with
above average luminosities will be over-represented). The magnitude of the
bias depends on the amount of scatter in the \LM\ relation and on the slope
of the mass function around the flux limit (that is, the mass limit
corresponding to the flux limit when converted to a luminosity at the
redshift of interest) of the sample. As the mass function steepens at
higher masses, surveys with higher mass limits will suffer a larger
bias. The bias can affect the slope, normalisation, evolution and scatter
of the scaling relations measured with the sample.

The heterogeneous nature of our sample, with clusters drawn from a variety
of different samples, makes it difficult to quantify the effects of the
Eddington bias on the scaling relations. To illustrate this, consider three
surveys (all based on \ROSAT\ data) that are represented (incompletely) in
our sample are the Brightest Cluster Survey \citep[BCS;][]{ebe98} covering
$z\la0.3$, the Massive Cluster Survey \citep[MACS;][]{ebe01b} at $0.3\la
z\la0.6$, and the 400 Square Degree survey \citep[400SD;][]{bur06} at
$0.3\la z\la1$. The flux limits of these surveys correspond to masses
(within \rf) of approximately $5\times10^{14}\Msol$ (BCS at $z=0.2$),
$7\times10^{14}\Msol$ (MACS at $z=0.45$), $3\times10^{14}\Msol$ (400SD at
$z=0.7$). The bias thus varies across our sample, and is lowest for the
deeper surveys. As a simple test of the effect of the bias on the measured
evolution, the evolution was fit to the $z>0.6$ clusters alone. These
clusters are generally drawn from deep surveys for which the bias is
lower. The best fitting evolution was $a_{LY}=2.0\pm0.1$, lower than the
value measured using the whole sample ($a_{LY}=2.2\pm0.1$), and approaching
the self-similar value of $1.8$. This indicates that stronger than
self-similar evolution measured for the whole sample (\textsection
\ref{s.evol}) is due in part to the effects of bias. However, it should be
noted that the evolution is measured with respect to a local relation,
which is itself affected by Eddington bias at some level. The low scatter
in the \LY\ relation means that the effects of bias should be fairly small,
but these will be quantified using statistically complete samples at high and low
redshifts in a future study.

\subsection{Comparison with other measurements of scatter in the \LM\ relation.}
The scatter we find in the \LMf\ relation ($\sigma_L=0.17-0.39$) is
significantly lower than that measured by RB02 who found a total
(statistical and intrinsic) scatter of $0.74$ in a sample
of $106$ nearby clusters (we refit their data to measure the intrinsic
scatter alone, and found $\sigma_L=0.63$). Several important differences exist
between that work and our study. RB02 used luminosities including the
entire core region, derived masses from cluster temperatures assuming
isothermality, and finally, derived the masses used for the scatter measurement
within \rt, rather than our \rf. Indeed, for our sample, we
find $\sigma_L=0.39$ when all of the core emission is included in our \Lx\
measurements (see Table \ref{t.yl}), suggesting that the remaining
difference from the RB02 is due to the scatter in the \MT\ relation used to
derive their masses. 

Other recent work has also shown that the scatter in the \LM\ relation is
lower than previously thought. \citet{rei06} and \citet{sta06} compared
theoretical mass functions with the X-ray luminosity function measured from
the RB02 data to determine the scatter in the \LM\ relation. The measured
scatter depends strongly on the assumed cosmology, through the dependence
of the mass function on $\sigma_8$ and $\Omega_m$. Using the best-fitting
cosmological values from the WMAP year 3 results \citep[$\sigma_8=0.74$,
$\Omega_m=0.238$;][]{spe06}, \citet{rei06} concluded that the scatter is
small (but non-zero). \citet{sta06} showed that assuming the WMAP year 1
cosmology \citep[$\sigma_8=0.9$, $\Omega_m=0.29$;][]{spe03} results in a
large scatter in the \LM\ relation and suggested a compromise model with
$\sigma_8=0.85$ and $\Omega_m=0.24$ that gives a scatter of $0.21$ in \Mt\
for a given \Lx. The measurement from our sample that is most directly
comparable to this value is the scatter in \Mf\ for \Lx\ measured with the
cluster cores included, $\sigma_M=0.21$. This is in agreement with the
\citet{sta06} ``compromise cosmology'' value and is within the $90\%$ upper
limit on the scatter set by \citet{rei06} using the WMAP year 3 cosmology
(estimated by converting their upper limit on the bias factor to a
scatter). Note that we do not suggest that the scatter measured with our data
can be used to distinguish between different cosmological models.

\section{Conclusions}
We have used a sample of $115$ clusters of $\Mf\ga10^{14}\Msol$ spanning
the redshift range $0.1<z<1.3$ to study the \LY\ relation, and by proxy,
the \LMf\ relation. We verified the low scatter in the \YM\ relation and
its self-similar evolution to $z=0.6$ for clusters in our sample using mass
estimates from the literature. Our main conclusions are as follows.
\begin{itemize}

\item A strong correlation exists between \Lx\ and \Yx.

\item The scatter in the \LY\ (and hence \LMf) relations is dominated by
cluster cores, with a secondary contribution from the less relaxed systems
at larger radii.

\item Once a sufficiently large core region is excluded ($0.15\rf$ or
$0.15E(z)^{-2/3}\Mpc$) the scatter in the \LY\ relation, and by proxy \LMf\
relation are low ($11\%$ and $17\%$ respectively).

\item The scatter remains reasonably small for survey-quality data, where \Lx\ is
estimated from soft band count rates, increasing to $19\%$ (\LY) and $21\%$
(\LMf).

\item The slope and normalisation of the \LY\ (and by proxy \LMf) relation are
insensitive to the merger status of the clusters, although the scatter can
be further reduced by considering only the most relaxed clusters. 

\item The shape of the intrinsic scatter distribution about the \LY\ relation is
well described by a lognormal function. 

\item The evolution of the \LY\ relation appears consistent with the
self-similar prediction, though our lack of knowledge of the details of the
scatter and Eddington bias limit the strength of this conclusion.

\item The inferred scatter in \LMf\ is much smaller than that found by
previous studies because of {\it i)} our improved mass estimates using \Yx, and
{\it ii)} our conservative exclusion of core emission. This is consistent
with recent results by \citet{rei06} and \citet{sta06}.

\item For high-redshift ($z>0.5$) clusters, there is no need to exclude the
core emission. The scatter for high-z clusters with cores included is the
same as that measured for the entire sample with cores excluded. This is
likely due to the absence of cool core clusters at high redshifts \citep{vik06c}.
\end{itemize}

These results suggest that a simple luminosity measurement can provide an
effective mass proxy for clusters with low-quality data (\ie\ insufficient
to measure \Yx\ directly) out to high redshifts. This has important
applications for testing cosmological models with current and future X-ray
cluster surveys.

\acknowledgments
We thank Christine Jones, Alexey Vikhlinin and Thomas Reiprich for useful
discussions. We are grateful to the anonymous referee for their useful
comments. BJM is supported by NASA through Chandra Postdoctoral
Fellowship Award Number PF4-50034 issued by the Chandra X-ray Observatory
Center, which is operated by the Smithsonian Astrophysical Observatory for
and on behalf of NASA under contract NAS8-03060.

\bibliographystyle{mn2e}
\bibliography{clusters}

\end{document}